\input harvmac

\input amssym
\input epsf


\newfam\frakfam
\font\teneufm=eufm10
\font\seveneufm=eufm7
\font\fiveeufm=eufm5
\textfont\frakfam=\teneufm
\scriptfont\frakfam=\seveneufm
\scriptscriptfont\frakfam=\fiveeufm


\def\bb{
\font\tenmsb=msbm10
\font\sevenmsb=msbm7
\font\fivemsb=msbm5
\textfont1=\tenmsb
\scriptfont1=\sevenmsb
\scriptscriptfont1=\fivemsb
}



\newfam\dsromfam
\font\tendsrom=dsrom10
\textfont\dsromfam=\tendsrom
\def\ds{\fam\dsromfam \tendsrom}


\newfam\mbffam
\font\tenmbf=cmmib10
\font\sevenmbf=cmmib7
\font\fivembf=cmmib5
\textfont\mbffam=\tenmbf
\scriptfont\mbffam=\sevenmbf
\scriptscriptfont\mbffam=\fivembf


\newfam\mbfcalfam
\font\tenmbfcal=cmbsy10
\font\sevenmbfcal=cmbsy7
\font\fivembfcal=cmbsy5
\textfont\mbfcalfam=\tenmbfcal
\scriptfont\mbfcalfam=\sevenmbfcal
\scriptscriptfont\mbfcalfam=\fivembfcal


\newfam\mscrfam
\font\tenmscr=rsfs10
\font\sevenmscr=rsfs7
\font\fivemscr=rsfs5
\textfont\mscrfam=\tenmscr
\scriptfont\mscrfam=\sevenmscr
\scriptscriptfont\mscrfam=\fivemscr




\def\tilde{\widetilde}
\def\t{\tilde}

\def\bar{\overline}
\def\b{\bar}
\def\bsq#1{{{\b{#1}}^{\lower 2.5pt\hbox{$\scriptstyle 2$}}}}
\def\bexp#1#2{{{\b{#1}}^{\lower 2.5pt\hbox{$\scriptstyle #2$}}}}
\def\dotexp#1#2{{{#1}^{\lower 2.5pt\hbox{$\scriptstyle #2$}}}}


\def\rt2{\sqrt{2}}
\def\half {{1 \over 2}}
\def\Re{\mathop{\rm Re}}
\def\Im{\mathop{\rm Im}}
\def\d{\partial}

\def\grad{\nabla}

\def\Tr{\mathop{\rm Tr}}


\font\tenbifull=cmmib10
\font\tenbimed=cmmib7
\font\tenbismall=cmmib5
\textfont9=\tenbifull \scriptfont9=\tenbimed
\scriptscriptfont9=\tenbismall

\mathchardef\bbGamma="7000
\mathchardef\bbDelta="7001
\mathchardef\bbPhi="7002
\mathchardef\bbAlpha="7003
\mathchardef\bbXi="7004
\mathchardef\bbPi="7005
\mathchardef\bbSigma="7006
\mathchardef\bbUpsilon="7007
\mathchardef\bbTheta="7008
\mathchardef\bbPsi="7009
\mathchardef\bbOmega="700A
\mathchardef\bbalpha="710B
\mathchardef\bbbeta="710C
\mathchardef\bbgamma="710D
\mathchardef\bbdelta="710E
\mathchardef\bbepsilon="710F
\mathchardef\bbzeta="7110
\mathchardef\bbeta="7111
\mathchardef\bbtheta="7112
\mathchardef\bbiota="7113
\mathchardef\bbkappa="7114
\mathchardef\bblambda="7115
\mathchardef\bbmu="7116
\mathchardef\bbnu="7117
\mathchardef\bbxi="7118
\mathchardef\bbpi="7119
\mathchardef\bbrho="711A
\mathchardef\bbsigma="711B
\mathchardef\bbtau="711C
\mathchardef\bbupsilon="711D
\mathchardef\bbphi="711E
\mathchardef\bbchi="711F
\mathchardef\bbpsi="7120
\mathchardef\bbomega="7121
\mathchardef\bbvarepsilon="7122
\mathchardef\bbvartheta="7123
\mathchardef\bbvarpi="7124
\mathchardef\bbvarrho="7125
\mathchardef\bbvarsigma="7126
\mathchardef\bbvarphi="7127




\def\thetasq{\theta^2}


\def\CH{{\cal H}}

\def\CJ{{\cal J}}

\def\CN{{\cal N}}
\def\CO{{\cal O}}

\def\CR{{\cal R}}

\def\CV{{\cal V}}


\def\1{{\ds 1}}

\def\Z{\hbox{$\bb Z$}}


\def\ep{\varepsilon}

\noblackbox

\def\unit{\relax{\rm 1\kern-.26em I}}
\def\nada{\relax{\rm 0\kern-.30em l}}
\def\tilde{\widetilde}
\def\t{\tilde}


\noblackbox
\def\IL{\relax{\rm I\kern-.18em L}}
\def\IH{\relax{\rm I\kern-.18em H}}
\def\IR{\relax{\rm I\kern-.18em R}}
\def\IC{\relax\hbox{$\inbar\kern-.3em{\rm C}$}}
\def\IZ{\relax\ifmmode\mathchoice
{\hbox{\cmss Z\kern-.4em Z}}{\hbox{\cmss Z\kern-.4em Z}} {\lower.9pt\hbox{\cmsss Z\kern-.4em Z}}
{\lower1.2pt\hbox{\cmsss Z\kern-.4em Z}}\else{\cmss Z\kern-.4em Z}\fi}

\def\CN {{\cal N}}
\def\CR {{\cal R}}

\def\CJ {{\cal J}}
\def\partialslash{\not{\hbox{\kern-2pt $\partial$}}}

\def\CV {{\cal V}}
\def\CO {{\cal O}}

\def\CH {{\cal H}}


\def\CN {{\cal N}}

\def\CO {{\cal O}}

\def\CV{{\cal V }}

\def\Tr{{\rm Tr}}

\font\manual=manfnt \def\dbend{\lower3.5pt\hbox{\manual\char127}}

\def\IZ{\relax\ifmmode\mathchoice
{\hbox{\cmss Z\kern-.4em Z}}{\hbox{\cmss Z\kern-.4em Z}} {\lower.9pt\hbox{\cmsss Z\kern-.4em Z}}
{\lower1.2pt\hbox{\cmsss Z\kern-.4em Z}}\else{\cmss Z\kern-.4em Z}\fi}
\def\half {{1\over 2}}

\def\bar{\overline}

\def\CH{{\cal H}}

\def\rt2{\sqrt{2}}
\def\irt2{{1\over\sqrt{2}}}

\def\t{\tilde}

\def\slashchar#1{\setbox0=\hbox{$#1$}           
   \dimen0=\wd0                                 
   \setbox1=\hbox{/} \dimen1=\wd1               
   \ifdim\dimen0>\dimen1                        
      \rlap{\hbox to \dimen0{\hfil/\hfil}}      
      #1                                        
   \else                                        
      \rlap{\hbox to \dimen1{\hfil$#1$\hfil}}   
      /                                         
   \fi}

\def\foursqr#1#2{{\vcenter{\vbox{
    \hrule height.#2pt
    \hbox{\vrule width.#2pt height#1pt \kern#1pt
    \vrule width.#2pt}
    \hrule height.#2pt
    \hrule height.#2pt
    \hbox{\vrule width.#2pt height#1pt \kern#1pt
    \vrule width.#2pt}
    \hrule height.#2pt
        \hrule height.#2pt
    \hbox{\vrule width.#2pt height#1pt \kern#1pt
    \vrule width.#2pt}
    \hrule height.#2pt
        \hrule height.#2pt
    \hbox{\vrule width.#2pt height#1pt \kern#1pt
    \vrule width.#2pt}
    \hrule height.#2pt}}}}
\def\psqr#1#2{{\vcenter{\vbox{\hrule height.#2pt
    \hbox{\vrule width.#2pt height#1pt \kern#1pt
    \vrule width.#2pt}
    \hrule height.#2pt \hrule height.#2pt
    \hbox{\vrule width.#2pt height#1pt \kern#1pt
    \vrule width.#2pt}
    \hrule height.#2pt}}}}
\def\sqr#1#2{{\vcenter{\vbox{\hrule height.#2pt
    \hbox{\vrule width.#2pt height#1pt \kern#1pt
    \vrule width.#2pt}
    \hrule height.#2pt}}}}

\def\figin{\epsfcheck\figin}\def\figins{\epsfcheck\figins}
\def\epsfcheck{\ifx\epsfbox\UnDeFiNeD
\message{(NO epsf.tex, FIGURES WILL BE IGNORED)}
\gdef\figin##1{\vskip2in}\gdef\figins##1{\hskip.5in}
\else\message{(FIGURES WILL BE INCLUDED)}%
\gdef\figin##1{##1}\gdef\figins##1{##1}\fi}
\def\DefWarn#1{}
\def\figinsert{\goodbreak\midinsert}
\def\ifig#1#2#3{\DefWarn#1\xdef#1{fig.~\the\figno}
\writedef{#1\leftbracket fig.\noexpand~\the\figno}%
\figinsert\figin{\centerline{#3}}\medskip\centerline{\vbox{\baselineskip12pt \advance\hsize by
-1truein\noindent\footnotefont{\bf Fig.~\the\figno:\ } \it#2}}
\bigskip\endinsert\global\advance\figno by1}

\lref\KapustinKZ{
  A.~Kapustin, B.~Willett and I.~Yaakov,
  ``Exact Results for Wilson Loops in Superconformal Chern-Simons Theories with
  Matter,''
  JHEP {\bf 1003}, 089 (2010)
  [arXiv:0909.4559 [hep-th]].
}

\lref\JafferisUN{
  D.~L.~Jafferis,
  ``The Exact Superconformal R-Symmetry Extremizes Z,''
  arXiv:1012.3210 [hep-th].
}

\lref\HamaEA{
  N.~Hama, K.~Hosomichi and S.~Lee,
  ``SUSY Gauge Theories on Squashed Three-Spheres,''
  JHEP {\bf 1105}, 014 (2011)
  [arXiv:1102.4716 [hep-th]].
}

\lref\DimofteJU{
  T.~Dimofte, D.~Gaiotto and S.~Gukov,
  ``Gauge Theories Labelled by Three-Manifolds,''
[arXiv:1108.4389 [hep-th]].
}

\lref\ImamuraWG{
  Y.~Imamura and D.~Yokoyama,
  ``$\CN=2$ supersymmetric theories on squashed three-sphere,''
  arXiv:1109.4734 [hep-th].
}

\lref\ImamuraUW{
  Y.~Imamura,
  ``Relation between the 4d superconformal index and the $S^3$ partition
  function,''
  JHEP {\bf 1109}, 133 (2011)
  [arXiv:1104.4482 [hep-th]].
}

\lref\KomargodskiPC{
  Z.~Komargodski and N.~Seiberg,
  ``Comments on the Fayet-Iliopoulos Term in Field Theory and Supergravity,''
  JHEP {\bf 0906}, 007 (2009)
  [arXiv:0904.1159 [hep-th]].
}

\lref\KomargodskiRB{
  Z.~Komargodski and N.~Seiberg,
  ``Comments on Supercurrent Multiplets, Supersymmetric Field Theories and
  Supergravity,''
  JHEP {\bf 1007}, 017 (2010)
  [arXiv:1002.2228 [hep-th]].
}

\lref\DumitrescuIU{
  T.~T.~Dumitrescu and N.~Seiberg,
  ``Supercurrents and Brane Currents in Diverse Dimensions,''
  JHEP {\bf 1107}, 095 (2011)
  [arXiv:1106.0031 [hep-th]].
}

\lref\IntriligatorJJ{
  K.~A.~Intriligator, B.~Wecht,
  ``The Exact Superconformal R Symmetry Maximizes a,''
Nucl.\ Phys.\  {\bf B667}, 183-200 (2003). [hep-th/0304128].
}

\lref\BarnesBM{
  E.~Barnes, E.~Gorbatov, K.~A.~Intriligator, M.~Sudano and J.~Wright,
  ``The exact superconformal R-symmetry minimizes $\tau_{RR}$,''
  Nucl.\ Phys.\  B {\bf 730}, 210 (2005)
  [arXiv:hep-th/0507137].
}
\lref\BarnesBW{
  E.~Barnes, E.~Gorbatov, K.~A.~Intriligator and J.~Wright,
  ``Current correlators and AdS/CFT geometry,''
Nucl.\ Phys.\ B {\bf 732}, 89 (2006).
[hep-th/0507146].
}

\lref\FreedmanTZ{
  D.~Z.~Freedman, S.~D.~Mathur, A.~Matusis and L.~Rastelli,
  ``Correlation functions in the CFT(d) / AdS(d+1) correspondence,''
Nucl.\ Phys.\ B {\bf 546}, 96 (1999).
[hep-th/9804058].
}

\lref\SeibergQD{
  N.~Seiberg,
  ``Modifying the Sum Over Topological Sectors and Constraints on Supergravity,''
JHEP {\bf 1007}, 070 (2010).
[arXiv:1005.0002 [hep-th]].
}

\lref\WittenYA{
  E.~Witten,
  ``SL(2,Z) action on three-dimensional conformal field theories with Abelian symmetry,''
In *Shifman, M. (ed.) et al.: From fields to strings, vol. 2* 1173-1200.
[hep-th/0307041].
}

\lref\BeniniQS{
  F.~Benini, C.~Closset and S.~Cremonesi,
  ``Chiral flavors and M2-branes at toric CY4 singularities,''
JHEP {\bf 1002}, 036 (2010).
[arXiv:0911.4127 [hep-th]].
}

\lref\MaldacenaSS{
  J.~M.~Maldacena, G.~W.~Moore and N.~Seiberg,
  ``D-brane charges in five-brane backgrounds,''
JHEP {\bf 0110}, 005 (2001).
[hep-th/0108152].
}

\lref\JafferisTH{
  D.~L.~Jafferis,
  ``Quantum corrections to N=2 Chern-Simons theories with flavor and their AdS(4) duals,''
[arXiv:0911.4324 [hep-th]].
}

\lref\SeibergPQ{
  N.~Seiberg,
  ``Electric - Magnetic Duality in Supersymmetric nonAbelian Gauge Theories,''
Nucl.\ Phys.\  {\bf B435}, 129-146 (1995). [hep-th/9411149].
}

\lref\KutasovIY{
  D.~Kutasov, A.~Parnachev, D.~A.~Sahakyan,
 ``Central charges and U(1)(R) symmetries in N=1 superYang-Mills,''
JHEP {\bf 0311}, 013 (2003). [hep-th/0308071].
}

\lref\GaiottoAK{
  D.~Gaiotto and E.~Witten,
  ``S-Duality of Boundary Conditions In N=4 Super Yang-Mills Theory,''
[arXiv:0807.3720 [hep-th]].
}

\lref\IntriligatorMI{
  K.~A.~Intriligator, B.~Wecht,
  ``RG fixed points and flows in SQCD with adjoints,''
Nucl.\ Phys.\  {\bf B677}, 223-272 (2004). [hep-th/0309201].
}

\lref\CardyCWA{
  J.~L.~Cardy,
  ``Is There a c Theorem in Four-Dimensions?,''
Phys.\ Lett.\  {\bf B215}, 749-752 (1988). }

\lref\ZamolodchikovGT{
  A.~B.~Zamolodchikov,
  ``Irreversibility of the Flux of the Renormalization Group in a 2D Field Theory,''
JETP Lett.\  {\bf 43}, 730-732 (1986). }

\lref\MyersTJ{
  R.~C.~Myers, A.~Sinha,
 ``Holographic c-Theorems in Arbitrary Dimensions,''
JHEP {\bf 1101}, 125 (2011).
[arXiv:1011.5819 [hep-th]].
}

\lref\JafferisZI{
  D.~L.~Jafferis, I.~R.~Klebanov, S.~S.~Pufu, B.~R.~Safdi,
  ``Towards the F-Theorem: N=2 Field Theories on the Three-Sphere,''
JHEP {\bf 1106}, 102 (2011).
[arXiv:1103.1181 [hep-th]].
}

\lref\KlebanovGS{
  I.~R.~Klebanov, S.~S.~Pufu, B.~R.~Safdi,
 ``F-Theorem without Supersymmetry,''
[arXiv:1105.4598 [hep-th]].
}

\lref\KomargodskiVJ{
  Z.~Komargodski, A.~Schwimmer,
  ``On Renormalization Group Flows in Four Dimensions,''
[arXiv:1107.3987 [hep-th]].
}

\lref\GreenDA{
  D.~Green, Z.~Komargodski, N.~Seiberg, Y.~Tachikawa and B.~Wecht,
 ``Exactly Marginal Deformations and Global Symmetries,''
  JHEP {\bf 1006}, 106 (2010)
  [arXiv:1005.3546 [hep-th]].
}

\lref\PolchinskiDY{
  J.~Polchinski,
  ``Scale and Conformal Invariance in Quantum Field Theory,''
  Nucl.\ Phys.\  B {\bf 303}, 226 (1988).
}

\lref\SchwimmerZA{
  A.~Schwimmer and S.~Theisen,
  ``Spontaneous Breaking of Conformal Invariance and Trace Anomaly Matching,''
  Nucl.\ Phys.\  B {\bf 847}, 590 (2011)
  [arXiv:1011.0696 [hep-th]].
}

\lref\JafferisNS{
  D.~Jafferis and X.~Yin,
  ``A Duality Appetizer,''
[arXiv:1103.5700 [hep-th]].
}

\lref\JafferisZI{
  D.~L.~Jafferis, I.~R.~Klebanov, S.~S.~Pufu and B.~R.~Safdi,
  ``Towards the F-Theorem: N=2 Field Theories on the Three-Sphere,''
  JHEP {\bf 1106}, 102 (2011)
  [arXiv:1103.1181 [hep-th]].
}

\lref\CasiniKV{
  H.~Casini, M.~Huerta and R.~C.~Myers,
  ``Towards a derivation of holographic entanglement entropy,''
JHEP\ {\bf 1105}, 036  (2011).
[arXiv:1102.0440 [hep-th]].
}

\lref\GreenDA{
  D.~Green, Z.~Komargodski, N.~Seiberg, Y.~Tachikawa and B.~Wecht,
  ``Exactly Marginal Deformations and Global Symmetries,''
  JHEP {\bf 1006}, 106 (2010)
  [arXiv:1005.3546 [hep-th]].
}

\lref\AlvarezGaumeIG{
  L.~Alvarez-Gaume and E.~Witten,
  ``Gravitational Anomalies,''
Nucl.\ Phys.\ B\ {\bf 234}, 269  (1984).
}

\lref\KomargodskiXV{
  Z.~Komargodski,
  ``The Constraints of Conformal Symmetry on RG Flows,''
  arXiv:1112.4538 [hep-th].
}

\lref\PestunRZ{
  V.~Pestun,
  ``Localization of gauge theory on a four-sphere and supersymmetric Wilson
  loops,''
  arXiv:0712.2824 [hep-th].
}

\lref\TachikawaTQ{
  Y.~Tachikawa,
  ``Five-dimensional supergravity dual of a-maximization,''
  Nucl.\ Phys.\  B {\bf 733}, 188 (2006)
  [arXiv:hep-th/0507057].
}

\lref\AmaritiXP{
  A.~Amariti and M.~Siani,
  ``F-maximization along the RG flows: a proposal,''
  JHEP {\bf 1111}, 056 (2011)
  [arXiv:1105.3979 [hep-th]].
}

\lref\MoritaCS{
  T.~Morita and V.~Niarchos,
  ``F-theorem, duality and SUSY breaking in one-adjoint Chern-Simons-Matter
  theories,''
  Nucl.\ Phys.\  B {\bf 858}, 84 (2012)
  [arXiv:1108.4963 [hep-th]].
}

\lref\KlebanovTD{
  I.~R.~Klebanov, S.~S.~Pufu, S.~Sachdev and B.~R.~Safdi,
  ``Entanglement Entropy of 3-d Conformal Gauge Theories with Many Flavors,''
  arXiv:1112.5342 [hep-th].
}

\lref\FestucciaWS{
  G.~Festuccia and N.~Seiberg,
  ``Rigid Supersymmetric Theories in Curved Superspace,''
JHEP {\bf 1106}, 114 (2011).
[arXiv:1105.0689 [hep-th]].
}

\lref\WessCP{
  J.~Wess, J.~Bagger,
  ``Supersymmetry and Supergravity,''
Princeton, Univ. Pr. (1992).
}

\lref\FerraraTX{
  S.~Ferrara, M.~Porrati,
  ``Central extensions of supersymmetry in four-dimensions and three-dimensions,''
Phys.\ Lett.\  {\bf B423}, 255-260 (1998).
[hep-th/9711116].
}

\lref\SohniusTP{
  M.~F.~Sohnius and P.~C.~West,
  ``An Alternative Minimal Off-Shell Version of N=1 Supergravity,''
Phys.\ Lett.\ B {\bf 105}, 353 (1981).
}

\lref\BeniniMF{
  F.~Benini, C.~Closset and S.~Cremonesi,
  ``Comments on 3d Seiberg-like dualities,''
  JHEP {\bf 1110}, 075 (2011)
  [arXiv:1108.5373 [hep-th]].
}

\lref\HamaAV{
  N.~Hama, K.~Hosomichi and S.~Lee,
  ``Notes on SUSY Gauge Theories on Three-Sphere,''
JHEP {\bf 1103}, 127 (2011).
[arXiv:1012.3512 [hep-th]].
}

\lref\MoritaCS{
  T.~Morita and V.~Niarchos,
  ``F-theorem, duality and SUSY breaking in one-adjoint Chern-Simons-Matter
  theories,''
  Nucl.\ Phys.\  B {\bf 858}, 84 (2012)
  [arXiv:1108.4963 [hep-th]].
}

\lref\FestucciaWS{
  G.~Festuccia and N.~Seiberg,
  ``Rigid Supersymmetric Theories in Curved Superspace,''
JHEP {\bf 1106}, 114 (2011).
[arXiv:1105.0689 [hep-th]].
}

\lref\CasiniKV{
  H.~Casini, M.~Huerta and R.~C.~Myers,
  ``Towards a derivation of holographic entanglement entropy,''
  JHEP {\bf 1105}, 036 (2011)
  [arXiv:1102.0440 [hep-th]].
}
\lref\BanksZN{
  T.~Banks and N.~Seiberg,
  ``Symmetries and Strings in Field Theory and Gravity,''
Phys.\ Rev.\ D {\bf 83}, 084019 (2011).
[arXiv:1011.5120 [hep-th]].
}

\lref\MinwallaMA{
  S.~Minwalla, P.~Narayan, T.~Sharma, V.~Umesh and X.~Yin,
  ``Supersymmetric States in Large N Chern-Simons-Matter Theories,''
  arXiv:1104.0680 [hep-th].
}

\lref\DeserSW{
  S.~Deser and J.~H.~Kay,
  ``Topologically Massive Supergravity,''
Phys.\ Lett.\ B {\bf 120}, 97 (1983).
}

\lref\SeibergVC{
  N.~Seiberg,
  ``Naturalness versus supersymmetric nonrenormalization theorems,''
Phys.\ Lett.\ B {\bf 318}, 469 (1993).
[hep-ph/9309335].
}

\lref\AharonyBX{
  O.~Aharony, A.~Hanany, K.~A.~Intriligator, N.~Seiberg and M.~J.~Strassler,
  ``Aspects of N = 2 supersymmetric gauge theories in three dimensions,''
  Nucl.\ Phys.\  B {\bf 499}, 67 (1997)
  [arXiv:hep-th/9703110].
}
\lref\AmaritiDA{
  A.~Amariti and M.~Siani,
  ``Z-extremization and F-theorem in Chern-Simons matter theories,''
  JHEP {\bf 1110}, 016 (2011)
  [arXiv:1105.0933 [hep-th]].
}

\lref\BelovZE{
  D.~Belov and G.~W.~Moore,
  ``Classification of Abelian spin Chern-Simons theories,''
[hep-th/0505235].
}
\lref\KapustinHK{
  A.~Kapustin and N.~Saulina,
  ``Topological boundary conditions in abelian Chern-Simons theory,''
Nucl.\ Phys.\ B {\bf 845}, 393 (2011).
[arXiv:1008.0654 [hep-th]].
}

\lref\KomargodskiRB{
  Z.~Komargodski and N.~Seiberg,
  ``Comments on Supercurrent Multiplets, Supersymmetric Field Theories and
  Supergravity,''
  JHEP {\bf 1007}, 017 (2010)
  [arXiv:1002.2228 [hep-th]].
}

\lref\MartelliTP{
  D.~Martelli, J.~Sparks and S.~T.~Yau,
  ``The geometric dual of a-maximisation for toric Sasaki-Einstein
  manifolds,''
  Commun.\ Math.\ Phys.\  {\bf 268}, 39 (2006)
  [arXiv:hep-th/0503183].
}

\lref\ButtiVN{
  A.~Butti and A.~Zaffaroni,
  ``R-charges from toric diagrams and the equivalence of a-maximization and
  Z-minimization,''
  JHEP {\bf 0511}, 019 (2005)
  [arXiv:hep-th/0506232].
}

\lref\BarnesBW{
  E.~Barnes, E.~Gorbatov, K.~A.~Intriligator and J.~Wright,
  ``Current correlators and AdS/CFT geometry,''
  Nucl.\ Phys.\  B {\bf 732}, 89 (2006)
  [arXiv:hep-th/0507146].
}

\lref\RedlichKN{
  A.~N.~Redlich,
  ``Gauge Non-Invariance and Parity Non-Conservation of Three-Dimensional
  Fermions,''
  Phys.\ Rev.\ Lett.\  {\bf 52}, 18 (1984).
}

\lref\RedlichDV{
  A.~N.~Redlich,
  ``Parity Violation And Gauge Noninvariance Of The Effective Gauge Field
  Action In Three-Dimensions,''
  Phys.\ Rev.\  D {\bf 29}, 2366 (1984).
}

\lref\AlvarezGaumeIG{
  L.~Alvarez-Gaume and E.~Witten,
  ``Gravitational Anomalies,''
  Nucl.\ Phys.\  B {\bf 234}, 269 (1984).
}

\lref\WittenHF{
  E.~Witten,
  ``Quantum field theory and the Jones polynomial,''
  Commun.\ Math.\ Phys.\  {\bf 121}, 351 (1989).
}

\lref\WittenZE{
  E.~Witten,
  ``Topological Quantum Field Theory,''
Commun.\ Math.\ Phys.\  {\bf 117}, 353 (1988).
}

\lref\CasiniEI{
  H.~Casini and M.~Huerta,
  ``On the RG running of the entanglement entropy of a circle,''
[arXiv:1202.5650 [hep-th]].
}

\lref\WittenYA{
  E.~Witten,
  ``SL(2,Z) action on three-dimensional conformal field theories with Abelian
  symmetry,''
  arXiv:hep-th/0307041.
}

\lref\GaiottoAK{
  D.~Gaiotto and E.~Witten,
  ``S-Duality of Boundary Conditions In N=4 Super Yang-Mills Theory,''
  arXiv:0807.3720 [hep-th].
}

\lref\GaddeIA{
  A.~Gadde and W.~Yan,
  ``Reducing the 4d Index to the $S^3$ Partition Function,''
  arXiv:1104.2592 [hep-th].
}

\lref\ColemanZI{
  S.~R.~Coleman and B.~R.~Hill,
  ``No More Corrections to the Topological Mass Term in QED in Three-Dimensions,''
Phys.\ Lett.\ B {\bf 159}, 184 (1985).
}

\lref\MartelliQJ{
  D.~Martelli and J.~Sparks,
  ``The large N limit of quiver matrix models and Sasaki-Einstein manifolds,''
Phys.\ Rev.\ D {\bf 84}, 046008 (2011).
[arXiv:1102.5289 [hep-th]].
}

\lref\MartelliYB{
  D.~Martelli, J.~Sparks and S.~-T.~Yau,
  ``Sasaki-Einstein manifolds and volume minimisation,''
Commun.\ Math.\ Phys.\  {\bf 280}, 611 (2008).
[hep-th/0603021].
}

\lref\EagerYU{
  R.~Eager,
  ``Equivalence of A-Maximization and Volume Minimization,''
[arXiv:1011.1809 [hep-th]].
}

\lref\MaldacenaRE{
  J.~M.~Maldacena,
  ``The Large N limit of superconformal field theories and supergravity,''
Adv.\ Theor.\ Math.\ Phys.\  {\bf 2}, 231 (1998), [Int.\ J.\ Theor.\ Phys.\  {\bf 38}, 1113 (1999)].
[hep-th/9711200].
}

\lref\GubserBC{
  S.~S.~Gubser, I.~R.~Klebanov and A.~M.~Polyakov,
  ``Gauge theory correlators from noncritical string theory,''
Phys.\ Lett.\ B {\bf 428}, 105 (1998).
[hep-th/9802109].
}

\lref\WittenQJ{
  E.~Witten,
  ``Anti-de Sitter space and holography,''
Adv.\ Theor.\ Math.\ Phys.\  {\bf 2}, 253 (1998).
[hep-th/9802150].
}

\lref\LiuEE{
  H.~Liu and M.~Mezei,
  ``A Refinement of entanglement entropy and the number of degrees of freedom,''
[arXiv:1202.2070 [hep-th]].
}

\lref\MyersED{
  R.~C.~Myers and A.~Singh,
  ``Comments on Holographic Entanglement Entropy and RG Flows,''
[arXiv:1202.2068 [hep-th]].
}

\lref\KapustinHA{
  A.~Kapustin and M.~J.~Strassler,
  ``On mirror symmetry in three-dimensional Abelian gauge theories,''
JHEP {\bf 9904}, 021 (1999).
[hep-th/9902033].
}

\lref\future{
  C.~Closset, T.~T.~Dumitrescu, G.~Festuccia, Z.~Komargodski and N.~Seiberg,
  ``Comments on Chern-Simons Contact Terms in Three Dimensions,''
[arXiv:1206.5218 [hep-th]].
}

\lref\KapustinXQ{
  A.~Kapustin, B.~Willett and I.~Yaakov,
  ``Nonperturbative Tests of Three-Dimensional Dualities,''
JHEP {\bf 1010}, 013 (2010).
[arXiv:1003.5694 [hep-th]].
}

\lref\KapustinMH{
  A.~Kapustin, B.~Willett and I.~Yaakov,
  ``Tests of Seiberg-like Duality in Three Dimensions,''
[arXiv:1012.4021 [hep-th]].
}

\lref\WillettGP{
  B.~Willett and I.~Yaakov,
  ``N=2 Dualities and Z Extremization in Three Dimensions,''
[arXiv:1104.0487 [hep-th]].
}

\lref\KutasovIY{
  D.~Kutasov, A.~Parnachev and D.~A.~Sahakyan,
  ``Central charges and U(1)R symmetries in N = 1 super Yang-Mills,''
  JHEP {\bf 0311}, 013 (2003)
  [arXiv:hep-th/0308071].
}

\lref\SohniusFW{
  M.~Sohnius and P.~C.~West,
  ``The Tensor Calculus And Matter Coupling Of The Alternative Minimal Auxiliary Field Formulation Of N=1 Supergravity,''
Nucl.\ Phys.\ B {\bf 198}, 493 (1982).
}

\lref\RocekBK{
  M.~Rocek and P.~van Nieuwenhuizen,
  ``$N \geq  2$ Supersymmetric Chern-Simons Terms As $d = 3$ Extended Conformal Supergravity,''
Class.\ Quant.\ Grav.\  {\bf 3}, 43 (1986).
}

\lref\DolanRP{
  F.~A.~H.~Dolan, V.~P.~Spiridonov and G.~S.~Vartanov,
  ``From 4d superconformal indices to 3d partition functions,''
Phys.\ Lett.\ B {\bf 704}, 234 (2011).
[arXiv:1104.1787 [hep-th]].
}

\lref\MartelliFU{
  D.~Martelli, A.~Passias and J.~Sparks,
  ``The Gravity dual of supersymmetric gauge theories on a squashed three-sphere,''
[arXiv:1110.6400 [hep-th]].
}

\lref\MartelliFW{
  D.~Martelli and J.~Sparks,
  ``The Nuts and Bolts of Supersymmetric Gauge Theories on Biaxially squashed Three-Spheres,''
[arXiv:1111.6930 [hep-th]].
}

\lref\ElvangST{
  H.~Elvang, D.~Z.~Freedman, L.~-Y.~Hung, M.~Kiermaier, R.~C.~Myers and S.~Theisen,
  ``On renormalization group flows and the a-theorem in 6d,''
[arXiv:1205.3994 [hep-th]].
}

\lref\deBoerKR{
  J.~de Boer, K.~Hori and Y.~Oz,
  ``Dynamics of N=2 supersymmetric gauge theories in three-dimensions,''
Nucl.\ Phys.\ B {\bf 500}, 163 (1997).
[hep-th/9703100].
}

\lref\AgarwalWD{
  P.~Agarwal, A.~Amariti and M.~Siani,
  ``Refined Checks and Exact Dualities in Three Dimensions,''
[arXiv:1205.6798 [hep-th]].
}

\lref\KuzenkoXG{
  S.~M.~Kuzenko, U.~Lindstrom and G.~Tartaglino-Mazzucchelli,
  ``Off-shell supergravity-matter couplings in three dimensions,''
JHEP {\bf 1103}, 120 (2011).
[arXiv:1101.4013 [hep-th]].
}

\lref\KuzenkoRD{
  S.~M.~Kuzenko and G.~Tartaglino-Mazzucchelli,
  ``Three-dimensional N=2 (AdS) supergravity and associated supercurrents,''
JHEP {\bf 1112}, 052 (2011).
[arXiv:1109.0496 [hep-th]].
}

\lref\AchucarroVZ{
  A.~Achucarro and P.~K.~Townsend,
  ``A Chern-Simons Action for Three-Dimensional anti-de-Sitter Supergravity Theories,''
Phys.\ Lett.\ B {\bf 180}, 89 (1986).
}

\lref\AchucarroGM{
  A.~Achucarro and P.~K.~Townsend,
  ``Extended Supergravitites in $d = 2+1$ as Chern-Simons Theories,''
Phys.\ Lett.\ B {\bf 229}, 383 (1989).
}

\lref\BuicanTY{
  M.~Buican,
  ``A Conjectured Bound on Accidental Symmetries,''
Phys.\ Rev.\ D {\bf 85}, 025020 (2012).
[arXiv:1109.3279 [hep-th]].
}



\rightline{PUPT-2407}
\rightline{WIS/05/12-Feb-DPPA}
\Title{
} {\vbox{\centerline{Contact Terms, Unitarity, and $F$-Maximization}
\vskip2pt
\centerline{in Three-Dimensional Superconformal Theories}}}

\centerline{Cyril Closset,$^1$ Thomas T. Dumitrescu,$^{2}$ Guido Festuccia,$^3$}
\centerline{Zohar Komargodski,$^{1,3}$ and Nathan Seiberg\hskip1pt $^{3}$}
\bigskip
\centerline{ $^{1}$ {\it Weizmann Institute of Science, Rehovot
76100, Israel}}
 \centerline{$^{2}$ {\it Department of Physics, Princeton University, Princeton, NJ 08544, USA}}
  \centerline{$^{3}${\it
Institute for Advanced Study, Princeton, NJ 08540, USA}}

\vskip40pt

\noindent We consider three-dimensional~$\CN=2$ superconformal field theories on a three-sphere and analyze their free energy~$F$ as a function of background gauge and supergravity fields. A crucial role is played by certain local terms in these background fields, including several Chern-Simons terms. The presence of these terms clarifies a number of subtle {properties of~$F$. This understanding allows us to prove the~$F$-maximization principle. It also explains why computing~$F$ via localization leads to a complex answer, even though we expect it to be real in unitary theories.  We discuss several corollaries of our results and comment on the relation to the~$F$-theorem.

\vskip35pt

\Date{May 2012}

\newsec{Introduction}

Any conformal field theory~(CFT) in~$d$ dimensions can be placed on the~$d$-sphere~$S^d$ in a canonical, conformally invariant way, by using the stereographic map from flat Euclidean space. It is natural to study the partition function~$Z_{S^d}$ of the CFT compactified on~$S^d$, or the associated free energy,
\eqn\fe{F_d = - \log Z_{S^d}~.}
Since the sphere is compact, $F_d$ does not suffer from infrared~(IR) ambiguities. However, it is generally divergent in the ultraviolet~(UV). For instance, it may contain power divergences,
\eqn\pdiv{F_d \sim \left(\Lambda r\right)^d + \cdots~,}
where~$r$ is the radius of the sphere and~$\Lambda$ is a~UV cutoff. (The ellipsis denotes less divergent terms.) These power divergences depend on~$r$ and are inconsistent with conformal invariance. They should be set to zero by a local counterterm. In the example~\pdiv\ the divergence can be canceled by adjusting the cosmological constant counterterm~$\int_{S^d} \sqrt g \, d^d x\,$.

What remains after all power divergences have been eliminated depends on whether the number of dimensions is even or odd. If~$d$ is even, the free energy contains a logarithmic term in the radius,
\eqn\logdiv{F_{d} \sim a \log \left(\Lambda r\right) +\left({\rm finite}\right)~,}
which cannot be canceled by a local, diffeomorphism invariant counterterm. It reflects the well-known trace anomaly. The coefficient~$a$ is an intrinsic observable of the CFT, while the finite part of~$F_d$ depends on the choice of UV cutoff.

If~$d$ is odd, there are no local trace anomalies and we remain with a pure number~$F_d$. In unitary theories~$F_d$ is real.\foot{Since our entire discussion is in Euclidean signature, we will not distinguish between unitarity and reflection positivity.} There are no diffeomorphism invariant counterterms that can affect the value of~$F_d$, and hence any UV cutoff that respects diffeomorphism invariance leads to the same answer. For this reason,~$F_d$ is an intrinsic observable of the CFT.

In two and four dimensions, it was shown~\refs{\ZamolodchikovGT\CardyCWA\KomargodskiVJ-\KomargodskiXV} that every unitary renormalization group~(RG) flow connecting a~CFT$_{\rm UV}$ at short distances to a~CFT$_{\rm IR}$ at long distances must respect the inequality
\eqn\athm{a_{\rm UV} > a_{\rm IR}~.}
See~\ElvangST\ for a discussion of the six-dimensional case. (Another quantity conjectured to decrease under RG flow was recently discussed in~\BuicanTY.) It has been proposed~\refs{\MyersTJ\CasiniKV\JafferisZI\MyersED-\LiuEE} that a similar inequality should hold in three dimensions,
\eqn\fthm{F_{\rm UV} > F_{\rm IR}~.}
(Since we will remain in three dimensions for the remainder of this paper, we have dropped the subscript~$d=3$.) This conjectured~$F$-theorem has been checked for a variety of supersymmetric flows, and also for some non-supersymmetric ones; see for instance~\refs{\MinwallaMA\AmaritiDA\AmaritiXP\KlebanovGS\MoritaCS\BeniniMF-\KlebanovTD}. Moreover, the free energy~$F$ on a three-sphere corresponds to a certain entanglement entropy~\CasiniKV.  This relation has been used recently~\CasiniEI\ to argue for~\fthm.

In practice, the free energy~$F$ is not easy to compute. Much recent work has focused on evaluating~$F$ in~$\CN=2$ superconformal theories (SCFTs). (The flat-space dynamics of~$\CN=2$ theories in three dimensions was first studied in~\refs{\deBoerKR,\AharonyBX}.) In such theories, it is possible to compute~$F$ exactly via localization~\WittenZE, which reduces the entire functional integral to a finite-dimensional matrix model~\refs{\KapustinKZ\JafferisUN-\HamaAV}. In this approach, one embeds the SCFT into the deep IR of a renormalization group flow from a free UV theory. The functional integral is then computed in this~UV description and reduces to an integral over a finite number of zero modes. (A similar reduction of the functional integral occurs in certain four-dimensional field theories~\PestunRZ.)

Since this procedure breaks conformal invariance, the theory can no longer be placed on the sphere in a canonical way. Nevertheless, it is possible to place the theory on~$S^3$ while preserving supersymmetry, and explicit Lagrangians were constructed in~\refs{\KapustinKZ\JafferisUN-\HamaAV}. A systematic approach to this subject was developed in~\FestucciaWS, where supersymmetric Lagrangians on curved manifolds were described in terms of background supergravity fields. This point of view will be important below. One finds that if the non-conformal theory has a~$U(1)_R$ symmetry, it is possible to place it on~$S^3$ while preserving an~$SU(2|1) \times SU(2)$ symmetry.  This superalgebra is a subalgebra of the superconformal algebra on the sphere, but as emphasized in~\FestucciaWS, its presence is not related to superconformal invariance.

The choice of~$SU(2|1) \times SU(2)$ symmetry is not unique. It depends on a continuous choice of~$R$-symmetry in the UV, as well as a discrete choice of orientation on the sphere.\foot{The orientation determines whether the bosonic~$SU(2) \subset SU(2|1)$ is the~$SU(2)_l$ or the~$SU(2)_r$ subgroup of the~$SU(2)_l \times SU(2)_r$ isometry group. Below, we will always assume the former.}  Given any reference~$R$-symmetry~$R_0$, the space of~$R$-symmetries is parameterized by the mixing with all Abelian flavor symmetries~$Q_a$,
\eqn\rmix{R(t) = R_0 + \sum_a t^a Q_a~.}
The free energy~$F(t)$ explicitly depends on the real parameters~$t^a$. Surprisingly, the function~$F(t)$ is complex-valued~\refs{\KapustinKZ\JafferisUN-\HamaAV}, even though we expect it to be real in a unitary theory. This will be discussed extensively below. In order to make contact with the free energy of the SCFT, we must find the values~$t^a = t_*^a$, such that~$R(t_*)$ is the~$R$-symmetry that appears in the~$\CN = 2$ superconformal algebra.

In this paper, we will show that the real part~$\Re F(t)$ satisfies
\eqn\fmax{{\d \over \d t^a} \Re F \biggr|_{t = t_*} = 0~,
\qquad {\d^2 \over \d t^a \d t^b} \Re F  \biggr|_{t = t_*} = -{\pi^2 \over 2} \, \tau_{a b}~.}
The matrix~$\tau_{a b}$ is determined by the flat-space two-point functions of the Abelian flavor currents~$j^\mu_a$ at separated points,
\eqn\currtp{\langle j^\mu_a (x) j_b^\nu (0)\rangle = {\tau_{a b} \over 16 \pi^2 } \left(\delta^{\mu\nu} \d^2 - \d^\mu \d^\nu\right) {1\over x^2}~.}
In a unitary theory~$\tau_{ab}$ is a positive definite matrix.

These conditions can be stated as a maximization principle: {\it the superconformal~$R$-symmetry~$R(t_*)$ locally maximizes~$\Re F(t)$ over the space of trial~$R$-symmetries~$R(t)$.} The local maximum~$\Re F(t_*)$ is the SCFT partition function on~$S^3$. This~$F$-maximization principle is similar to~$a$-maximization in four dimensions~\IntriligatorJJ.
Analogously, it leads to~\fthm\ for a wide variety of renormalization group flows. The first condition in~\fmax\ is the extremization condition proposed in~\JafferisUN. The fact that the extremum should be a maximum was conjectured in~\JafferisZI.

A corollary of~\fmax\ is that~$\tau_{ab}$ is constant on conformal manifolds. It does not depend on deformations of the SCFT by exactly marginal operators, as long as these operators do not break the associated flavor symmetries.  Another consequence of~\fmax\ is that~$\tau_{ab}$ can be obtained from the same matrix integral that calculates the free energy, adding to the list of SCFT observables that can be computed exactly using localization. Below, we will discuss several new observables that can also be extracted from~$F(t)$.

We will establish~\fmax\ by studying the free energy of the SCFT as a function of background gauge fields for the flavor currents~$j^\mu_a$, as well as various background supergravity fields.  In theories with~$\CN=2$ supersymmetry, every flavor current is embedded in a real linear superfield~$\CJ_a$ and the corresponding background gauge field resides in a vector superfield~$\CV^a$. The supergravity fields are embedded in a multiplet $\CH$. The free energy~$F[\CV^a,\CH]$ of the SCFT now depends on these sources.

Localization allows us to compute~$F[\CV^a,\CH]$ for certain special values of the background fields~$\CV^a, \CH$. On a three-sphere, the answer turns out to violate several physical requirements: it is not real, in contradiction with unitarity, and it is not conformally invariant. The imaginary part arises because we must assign imaginary values to some of the background fields in order to preserve rigid supersymmetry on the sphere~\FestucciaWS. The lack of conformal invariance is more subtle. It reflects a new anomaly in three-dimensional~$\CN=2$ superconformal theories~\future.

As we will see below, $F[\CV^a,\CH]$ may contain Chern-Simons terms in the background fields, which capture contact terms in correlation functions of various currents. For instance, a contact term
\eqn\flavct{\langle j_a^\mu(x) j_b^\nu(0)\rangle = \cdots + {i \kappa_{ab} \over 2\pi} \ep^{\mu\nu\rho} \d_\rho \delta^{(3)}(x)~,}
corresponds to a Chern-Simons term for the background gauge fields~$\CV^a$ and~$\CV^b$. Such contact terms are thoroughly discussed in~\future, where it is shown that they lead to new observables in three-dimensional conformal field theories. Here we will use them to elucidate various properties of the three-sphere partition function in~$\CN=2$ superconformal theories. In particular, we explain why some of these terms are responsible for the fact that~$F[\CV^a, \CH]$ is not conformally invariant. Moreover, we show how the observables related to~$\kappa_{ab}$ in~\flavct\ can be computed exactly using localization.\foot{In this paper we explain how to compute the quantities~$
\tau_{ab}$ and~$\kappa_{ab}$, which are associated with global flavor symmetries, using localization. The corresponding observables for the~$R$-symmetry, and other closely related objects, can also be computed exactly. We leave a detailed discussion to future work.}

The outline of this paper is as follows. Section~2 summarizes necessary material from~\future. We introduce the background fields~$\CV^a, ~\CH$ and present various Chern-Simons terms in these fields. We explain why they give rise to new observables and how some of them lead to a violation of conformal invariance.
In section~3 we place the theory on a three-sphere and review the relevant supergravity background that leads to rigid supersymmetry~\FestucciaWS. We then relate the linear and quadratic terms in the background gauge fields~$\CV^a$ to the flat-space quantities introduced in section~2.
In section~4 we derive~\fmax\ and clarify the relation to~\fthm. Section~5 contains some simple examples.

\newsec{Background Fields and Contact Terms}

In this section we discuss contact terms in two-point functions of conserved currents. In theories with~$\CN=2$ supersymmetry, we distinguish between~$U(1)$ flavor currents and~$U(1)_R$ currents. These contact terms correspond to Chern-Simons terms in background gauge and supergravity fields. Their fractional parts are meaningful physical observables and some of them lead to a new anomaly in~$\CN=2$ superconformal theories. This section is a summary of~\future.

\subsec{Non-Supersymmetric Theories}

Consider a three-dimensional conformal field theory with a global, compact~$U(1)$ symmetry, and the associated current~$j_\mu$. We can couple it to a background gauge field~$a_\mu$, and consider the free energy~$F[a]$, which is defined by
\eqn\fadef{e^{-F[a]} = \bigg< \exp\Big(\int d^3 x \, j_\mu a^\mu + \cdots \Big)\bigg>~.}
Here the ellipsis denotes higher-order terms in~$a_\mu$ that may be required in order to ensure invariance of~$F[a]$ under background gauge transformations of~$a_\mu$. A familiar example is the seagull term~$a_\mu a^\mu |\phi|^2$, which is needed when a charged scalar field~$\phi$ is coupled to~$a_\mu$.

We see from~\fadef\ that~$F[a]$ is the generating functional for connected correlation functions of~$j_\mu$. The two-point function~$\langle j_\mu(x) j_\nu(0)\rangle$ is constrained by current conservation and conformal symmetry, so that
\eqn\tpf{\langle j_\mu(x) j_\nu(0) \rangle = {\tau \over 16\pi^2}\left(\del^2\delta_{\mu\nu}-\del_\mu\del_\nu\right) {1\over x^2} + {i \kappa \over 2 \pi} \ep_{\mu\nu\rho} \d^\rho \delta^{(3)} (x)~.}
Here~$\tau$ and~$\kappa$ are dimensionless real constants. At separated points, only the first term contributes, and unitarity implies that~$\tau \geq 0$. (If~$\tau = 0$, then~$j_\mu$ is a redundant operator.) The correlation function at separated points gives rise to a non-local term in~$F[a]$. The term proportional to~$\kappa$ is a contact term, whose sign is not constrained by unitarity. It corresponds to a background Chern-Simons term in~$F[a]$,
\eqn\bcs{{i \kappa \over 4 \pi} \int d^3 x \, \ep^{\mu\nu\rho} a_\mu \d_\nu a_\rho~.}
This term explicitly breaks parity.

Correlation functions at separated points are universal.  They do not depend on short-distance physics. By contrast, contact terms depend on the choice of UV cutoff. They can be changed by adjusting local terms in the dynamical or background fields. Some contact terms are determined by imposing symmetries. For instance, the seagull term discussed above ensures current conservation. The contact term proportional to~$\kappa$ in~\tpf\ is not of this type. Nevertheless, it possesses certain universality properties, as we will now review.

The Chern-Simons term~\bcs\ is invariant under small background gauge transformations, as required by current conservation. However, it is not the integral of a gauge-invariant local density and this restricts the freedom in changing~$\kappa$ by adding a local counterterm in the exponent of~\fadef. This restriction arises because we can place the theory on a curved manifold that allows non-trivial bundles for the background gauge field~$a_\mu$. Demanding invariance under large gauge transformations implies that~$\kappa$ can only be shifted by an integer.\foot{Here we follow the common practice of attributing the quantization of Chern-Simons levels to invariance under large gauge transformations.  A more careful treatment involves a definition of the Chern-Simons term~\bcs\ using an extension of the gauge field~$a_\mu$ to an auxiliary four-manifold. Demanding that the answer be independent of how we choose this four-manifold leads to the same quantization condition as above.} Therefore, the fractional part of~$\kappa$ is universal and does not depend on the short-distance physics. It is an intrinsic observable of the CFT. If we choose to set~$\kappa$ to zero by a local counterterm, then~$F[a]$ is no longer invariant under large background gauge transformations: its imaginary part shifts by an amount that is determined by the observable described above and the topology of the gauge bundle.

As described in~\future, there are different ways to calculate this observable in flat space. Below, we will discuss its importance for supersymmetric theories on a three-sphere.

\subsec{Supersymmetric Theories}

In theories with~$\CN=2$ supersymmetry, we distinguish between~$U(1)$ flavor symmetries and~$U(1)_R$ symmetries. A global~$U(1)$ flavor current~$j_\mu$ is embedded in a real linear superfield~$\CJ$, which satisfies~$D^2 \CJ = \b D^2 \CJ = 0$.\foot{We follow the conventions of~\DumitrescuIU, continued to Euclidean signature. The gamma matrices are given by~${\left(\gamma^\mu\right)_\alpha}^\beta = \left(\sigma^3, -\sigma^1, -\sigma^2\right)$, where~$\sigma^i$ are the Pauli matrices. The totally antisymmetric Levi-Civita symbol is normalized so that~$\ep_{123} = 1$. Note the identity~$\gamma_\mu \gamma_\nu = \delta_{\mu\nu} + i \ep_{\mu\nu\rho} \gamma^\rho$.} In components,
\eqn\jcomp{\CJ = J + i \theta j + i \b \theta \b j + i \theta \b \theta K - \left(\theta \gamma^\mu \b \theta\right) j_\mu + \cdots~.}
Superconformal invariance implies that~$J, K$, and~$j_\mu$ are conformal primaries of dimension~$\Delta_J = 1$, $\Delta_K = 2$, and~$\Delta_{j_\mu} = 2$. (Only~$J$ is a superconformal primary.) It follows that the one-point functions of~$J$ and~$K$ vanish, while their two-point functions are related to the two-point function~\tpf\ of~$j_\mu$ with~$\tau = \tau_{ff}$ and~$\kappa = \kappa_{ff}$,
\eqn\threefun{\eqalign{& \langle J(x) J(0) \rangle = {\tau_{ff} \over 16\pi^2} {1\over x^2}~,\cr
& \langle K(x) K(0)\rangle = {\tau_{ff} \over 8\pi^2} {1\over x^4}~,\cr
&  \langle J(x) K(0)\rangle =  {\kappa_{ff} \over 2 \pi}  \delta^{(3)}(x)~.}}
The subscript~$ff$ emphasizes the fact that we are considering the two-point function of a flavor current. The constant~$\tau_{ff}$ is normalized so that~$\tau_{ff} = 1$ for a free chiral superfield of charge~$+1$.

We can couple~$\CJ$ to a background vector superfield,
\eqn\gaugemult{\CV = \cdots + \left(\theta \gamma^\mu \b \theta\right) a_\mu - i \theta \b \theta \sigma - i \thetasq \b \theta \b \lambda + i \b \thetasq \theta \lambda - \half \thetasq \b \thetasq D~.}
Here~$a_\mu, \sigma$, and~$D$ are real. Background gauge transformations shift~$\CV \rightarrow \CV + \Omega + \b \Omega$ with chiral~$\Omega$, so that~$\sigma$ and~$D$ are gauge invariant, while~$a_\mu$ transforms like an ordinary gauge field. (The ellipsis denotes fields that are pure gauge modes and do not appear in gauge-invariant functionals of~$\CV$.) The coupling of~$\CJ$ to~$\CV$ takes the form
\eqn\flatjv{2 \int d^4 \theta \, \CJ \CV = J D+ j_\mu a^\mu + K\sigma + \left({\rm fermions}\right)~.}
Now the free energy~$F[\CV]$ is a supersymmetric functional of the background gauge superfield~$\CV$. The supersymmetric generalization of the Chern-Simons term~\bcs\ takes the form
\eqn\ffcs{F_{ff} = -{\kappa_{ff} \over 2 \pi} \int d^3 x \int d^4 \theta \, \Sigma \CV = {\kappa_{ff} \over 4 \pi} \int d^3x \, \left(i \ep^{\mu\nu\rho} a_\mu \d_\nu a_\rho - 2 \sigma D\right) + \left({\rm fermions}\right)~.}
Here the real linear superfield~$\Sigma = {i \over 2} \b D D \CV$ is the gauge-invariant field strength corresponding to~$\CV$. This Chern-Simons term captures the contact terms in the two-point functions~\tpf\ and~\threefun.  It is conformally invariant.

A~$U(1)_R$ current~$j_\mu^{(R)}$ is embedded in a supercurrent multiplet~$\CR_\mu$, which also contains the supersymmetry current~$S_{\mu\alpha}$, the energy-momentum tensor~$T_{\mu\nu}$, a current~$j_\mu^{(Z)}$ that corresponds to the central charge~$Z$ in the supersymmetry algebra, and a string current~$\ep_{\mu\nu\rho} \d^\rho J^{(Z)}$. All of these currents are conserved. See~\DumitrescuIU\ for a thorough discussion of supercurrents in three dimensions. In components,
\eqn\rmultcomp{\eqalign{& \CR_\mu = j^{(R)}_\mu - i \theta S_\mu - i \b \theta \b S_\mu - \left(\theta \gamma^\nu \b \theta\right) \left(2 T_{\mu\nu}+ i\ep_{\mu\nu\rho} \d^\rho J^{(Z)}\right) \cr
& \hskip26pt - i \theta \b \theta \left(2 j^{(Z)}_\mu+ i \ep_{\mu\nu\rho} \d^\nu j^{(R) \rho}\right) + \cdots~.}}
Note that there are additional factors of~$i$ in~\rmultcomp\ compared to the formulas in~\DumitrescuIU, because we are working in Euclidean signature. (In Lorentzian signature the superfield~$\CR_\mu$ is real.)

The~$\CR$-multiplet couples to the linearized metric superfield~$\CH_\mu$. In Wess-Zumino gauge,
\eqn\hcomp{\CH_\mu = \half \left(\theta \gamma^\nu \b \theta\right) \left(h_{\mu\nu} - i B_{\mu\nu}\right) - {1 \over 2} \theta \b \theta C_\mu- {i \over 2} \thetasq \b \theta \b \psi_\mu +{i \over 2} \b \thetasq \theta \psi_\mu + \half \thetasq \b \thetasq \left(A_\mu-V_\mu\right)~.}
Here~$h_{\mu\nu}$ is the linearized metric, so that~$g_{\mu\nu} = \delta_{\mu\nu} + 2 h_{\mu\nu}$, and~$\psi_{\mu\alpha}$ is the gravitino. The vectors~$C_\mu$ and~$A_\mu$ are Abelian gauge fields, and~$B_{\mu\nu}$ is a two-form gauge field. We will also need the following field strengths,
\eqn\fsdef{\eqalign{& V_\mu = - \ep_{\mu\nu\rho} \d^\nu C^\rho~, \qquad \d^\mu V_\mu = 0~,\cr
& H = {1 \over 2} \ep_{\mu\nu\rho} \d^\mu B^{\nu\rho}~.}}
As above, there are several unfamiliar factors of~$i$ in~\hcomp\ that arise in Euclidean signature. The coupling of~$\CR_\mu$ to~$\CH_\mu$ takes the form
\eqn\rcoup{2 \int d^4 \theta \,  \CR_\mu \CH^\mu = T_{\mu\nu} h^{\mu\nu} - j^{(R)}_\mu \big(A^\mu - {3 \over 2} V^\mu\big) + i j^{(Z)}_\mu C^\mu - J^{(Z)} H + \left({\rm fermions}\right)~.}
Since the gauge field~$A_\mu$ couples to the~$R$-current, we see that the gauge freedom includes local~$R$-transformations. This is analogous to~$\CN=1$ new minimal supergravity in four dimensions~\refs{\SohniusTP,\SohniusFW}. For a recent discussion, see~\refs{\KuzenkoXG,\KuzenkoRD}.

If the theory is superconformal, the~$\CR$-multiplet reduces to a smaller supercurrent. Consequently, the linearized metric superfield~$\CH_\mu$ enjoys more gauge freedom, which allows us to set~$B_{\mu\nu}$ and~$A_\mu - \half V_\mu$ to zero. The combination~$A_\mu-{3\over 2}V_\mu$ remains and transforms like an Abelian gauge field.

Using~$\CH_\mu$, we can construct three Chern-Simons terms. They are derived in~\future. Surprisingly, not all of them are conformally invariant.\foot{In order to write suitably covariant formulas, we will include some terms that go beyond linearized supergravity, such as the measure factor~$\sqrt g$. We also endow~$\ep_{\mu\nu\rho}$ with a factor of~$\sqrt g$, so that it transforms like a tensor. Consequently, the field strength~$V_\mu = - \ep_{\mu\nu\rho} \d^\nu C^\rho$ is covariantly conserved,~$\grad_\mu V^\mu =0$.}

\medskip

\item{$\bullet$} {\it Gravitational Chern-Simons Term:}
\eqn\lcs{\eqalign{F_g = {\kappa_g \over 192 \pi} \int  \sqrt g \, d^3 x \,  \bigg(& i \ep^{\mu\nu\rho} \Tr \big(\omega_\mu \d_\nu \omega_\rho + {2 \over 3} \omega_\mu \omega_\nu \omega_\rho\big) \cr
& +4 i \ep^{\mu\nu\rho} \big(A_\mu - {3 \over 2} V_\mu\big) \d_\nu \big(A_\rho - {3 \over 2} V_\rho\big)\bigg) + \left({\rm fermions}\right)~.}}
Here~$\omega_\mu$ is the spin connection. We see that the~$\CN=2$ completion of the usual gravitational Chern-Simons term also involves a Chern-Simons term for~$A_\mu -{3 \over 2} V_\mu$. Like the flavor-flavor term, the gravitational Chern-Simons term is conformally invariant. It was previously studied in the context of  $\CN=2$ conformal supergravity~\RocekBK, see also~\refs{\AchucarroVZ,\AchucarroGM}.

\medskip

\item{$\bullet$} {\it $Z$-$Z$ Chern-Simons Term:}
\eqn\ggcs{F_{zz} = -{\kappa_{zz} \over 4 \pi} \int \sqrt g \, d^3 x \,  \left(i \ep^{\mu\nu\rho} \big(A_\mu - \half V_\mu \big) \d_\nu \big( A_\rho - \half V_\rho\big) + \half H R + \cdots \right) + \left({\rm fermions}\right)~.}
Here~$R$ is the Ricci scalar.\foot{In our conventions, a~$d$-dimensional sphere of radius~$r$ has scalar curvature~$R = -{d(d-1) \over r^2}$~.} The ellipsis denotes higher-order terms in the bosonic fields, which go beyond linearized supergravity. The $Z$-$Z$ Chern-Simons term is not conformally invariant, as is clear from the presence of the Ricci scalar.  This lack of conformal invariance is related to the following fact: in a superconformal theory, the~$\CR$-multiplet reduces to a smaller supercurrent and the operators conjugate to~$R$,~$H$ and $A_\mu-{1\over 2} V_\mu$ are redundant.

\medskip

\item{$\bullet$} {\it Flavor-$R$ Chern-Simons Term:}
\eqn\fgcs{F_{fr} = -{\kappa_{fr} \over 2 \pi} \int \sqrt g \, d^3 x \, \left(i \ep^{\mu\nu\rho} a_\mu\d_\nu \big(A_\rho - \half V_\rho\big) + {1 \over 4} \sigma R - D H + \cdots \right) + \left({\rm fermions}\right)~.}
The meaning of the ellipsis is as in~\ggcs\ above.  Again, the presence of~$R$,~$H$, and~$A_\mu - {1\over 2} V_\mu$ shows that this term is not conformally invariant. The relative sign between the Chern-Simons terms~\ffcs\ and~\fgcs\ is due to the different couplings~\flatjv\ and~\rcoup\ of~$j_\mu$ and~$j_\mu^{(R)}$ to their respective background gauge fields. Unlike the conformal Chern Simons terms~\ffcs\ and~\lcs, the $Z$-$Z$ term~\ggcs\ and the flavor-$R$ term~\fgcs\ are novel. Their lack of conformal invariance will be important below.

\bigskip

The Chern-Simons terms~\ffcs, \lcs, \ggcs, and~\fgcs\ summarize contact terms in two-point functions of~$\CJ$ and~$\CR_\mu$. As we stated above, the fractional parts of these contact terms are meaningful physical observables. This is thoroughly explained in~\future. Using the background fields~$\CV$ and~$\CH_\mu$, we can construct two additional local terms: the Fayet-Iliopoulos~(FI) term,
\eqn\fiterm{F_{FI} = \Lambda \int \sqrt g \, d^3x \left(D + \cdots\right) + \left({\rm fermions}\right)~,}
and the Einstein-Hilbert term,
\eqn\EH{F_{EH} = \Lambda \int \sqrt g \, d^3 x \,  \left(R + \cdots \right) + \left({\rm fermions}\right)~.}
These terms are not conformally invariant, and they are multiplied by an explicit power of the UV cutoff~$\Lambda$. They correspond to conventional contact terms, which can be adjusted at will. Below we will use them to remove certain linear divergences. A finite coefficient of~\fiterm\ leads to a one-point function for~$J$. In a scale-invariant theory it is natural to set such a dimensionful finite coefficient to zero.  More generally, the dynamical generation of FI-terms is very constrained. For a recent discussion, see~\refs{\KomargodskiPC,\KomargodskiRB,\DumitrescuIU} and references therein. Note that a cosmological constant counterterm proportional to~$\Lambda^3$ is not allowed by supersymmetry.

\subsec{A Superconformal Anomaly}

As we have seen above, the two Chern-Simons terms~\ggcs\ and~\fgcs\ are not conformally invariant. Moreover, we have argued that the fractional parts of their coefficients~$\kappa_{zz}$ and~$\kappa_{fr}$ are meaningful physical observables. If these fractional parts are non-vanishing, certain correlation functions have non-conformal contact terms. If we want to preserve supersymmetry, we have to choose between the following:
\medskip
\item{1.)} Retain these Chern-Simons terms at the expense of conformal invariance. In this case, the free energy is invariant under large background gauge transformations.
\item{2.)} Restore conformal invariance by adding appropriate Chern-Simons counterterms with fractional coefficients. In this case the free energy in the presence of topologically nontrivial background fields is not invariant under large gauge transformations. Its imaginary part, which encodes the fractional parts of~$\kappa_{zz}$ and~$\kappa_{fr}$, is only well defined if we specify additional geometric data. This is similar to the framing anomaly of~\WittenHF.
\medskip
\noindent
This understanding is essential for our discussion below. A detailed explanation can be found in~\future.
The second option above is the less radical of the two (the idea of adding Chern-Simons terms to a theory in order to ensure some physical requirements has already appeared long ago in several contexts~\refs{\RedlichKN\RedlichDV-\AlvarezGaumeIG,\WittenHF}), but we will explore both alternatives.

\newsec{The Free Energy on a Three-Sphere}

Coupling the flat-space theory to the background supergravity multiplet~$\CH$ renders it invariant under all background supergravity transformations. For certain expectation values of the fields in~$\CH$, the theory also preserves some amount of rigid supersymmetry~\FestucciaWS. Here we are interested in round spheres~\refs{\KapustinKZ\JafferisUN-\HamaAV,\FestucciaWS}.\foot{Recently, it was found that various squashed spheres also admit rigid supersymmetry~\refs{\HamaEA\DolanRP\GaddeIA\ImamuraUW \ImamuraWG\MartelliFU-\MartelliFW}. Many of our results can be generalized to these backgrounds.} In stereographic coordinates, the metric takes the form
\eqn\metric{g_{\mu\nu} = {4r^4  \over (r^2+x^2)^2} \, \delta_{\mu\nu}~,}
where~$r$ is the radius of the sphere. In order to preserve supersymmetry, we must also turn on a particular {\it imaginary} value for the background~$H$-flux \FestucciaWS,
\eqn\sphereflux{H = -{i \over r}~.}
This expectation value explicitly violates unitarity, since~$H$ is real in a unitary theory. Given a generic~$\CN=2$ theory with a choice of~$R$-symmetry, the background fields~\metric\ and~\sphereflux\ preserve an~$SU(2|1) \times SU(2)$ superalgebra. If the theory is superconformal, this is enhanced to the full superconformal algebra and the coupling to the background fields in~$\CH$ reduces to the one obtained by the stereographic map from flat space.  In this case the imaginary value for~$H$ in~\sphereflux\ is harmless and does not lead to any violations of unitarity~\FestucciaWS.

In this section, we will study an~$\CN = 2$ SCFT on a three-sphere and consider its free energy~$F[\CV]$ in the presence of a background gauge field~$\CV$ for the current~$\CJ$. For our purposes, it is sufficient to analyze~$F[\CV]$ for constant values of the background fields~$D$ and~$\sigma$. The other fields in~$\CV$ are set to zero. We will study~$F[\CV]$ as a power series expansion in~$D$ and~$\sigma$ around zero, starting with the free energy~$F[0]$ itself.

As we saw in the previous section, superconformal invariance may be violated by certain Chern-Simons contact terms. We can restore it by adding bare Chern-Simons counterterms with appropriate fractional coefficients, but this forces us to give up on invariance under large background gauge transformations. Here we will choose to retain the non-conformal terms and preserve invariance under large gauge transformations, since this setup is natural in calculations based on localization. Only the~$Z$-$Z$ Chern-Simons term~\ggcs\ and the gravitational Chern-Simons term~\lcs\ can contribute to~$F[0]$. On the sphere, the imaginary value of~$H$ in~\sphereflux\ implies that~$F_{zz}$ reduces to a purely imaginary constant, since the coefficient~$\kappa_{zz}$ in~\ggcs\ is real. The value of this constant depends on non-linear terms in the gravity fields, which are not captured by the linearized formula~\ggcs. The gravitational Chern-Simons term is superconformal and it does not contribute on the round sphere. In general, we will therefore find a complex~$F[0]$. Its real part is the conventional free energy of the SCFT, which must be real by unitarity. The imaginary part is due to a Chern-Simons term in the supergravity background fields.

The terms linear in~$D$ and~$\sigma$ reflect the one-point functions of~$J$ and~$K$. If our theory were fully conformally invariant, these terms would be absent.  However, in the presence of the non-conformal flavor-$R$ Chern-Simons term~\fgcs\ this is not the case. On the sphere, this term reduces to
\eqn\flavorgravity{
F_{fr} = {\kappa_{fr} \over 2 \pi}\int_{S^3} \sqrt g \, d^3x \,  \left({\sigma\over r^2}-{i D\over r}\right)~.
}
The explicit factor of~$i$, which violates unitarity, is due to the imaginary value of~$H$ in~\sphereflux. The relative coefficient between~$\sigma$ and~$D$ depends on both the linearized terms that appear explicitly in~\fgcs\ and on non-linear terms, denoted by an ellipsis.  Instead of computing them, we can check that~\flavorgravity\ is supersymmetric on the sphere. This term leads to non-trivial one-point functions for~$J$ and~$K$. However, the fact that~$\kappa_{fr}$ is real implies that
\eqn\Foneri{\partial_\sigma \Im F\bigl |_{\CV=0}=0~,\qquad \partial_D \Re F\bigl |_{\CV=0}=0~.}

In order to understand the terms quadratic in~$D$ and~$\sigma$, we must determine the two-point functions of $J$ and $K$ on the sphere. At separated points, they are easily obtained from the flat-space correlators~\threefun\ using the stereographic map,
\eqn\jjsthree{\eqalign{& \langle J(x) J(y)\rangle_{S^3}={\tau_{ff} \over 16\pi^2} {1 \over s(x,y)^{2}}~,\cr
& \langle K(x) K(y)\rangle_{S^3}={\tau_{ff} \over 8\pi^2} {1 \over s(x,y)^{4}}~,\cr
& \langle J(x) K(y)\rangle_{S^3} = 0~.}}
Here~$s(x, y)$ is the $SO(4)$ invariant distance function on the sphere. In stereographic coordinates,
\eqn\sfun{s(x,y) = {2r^2|x-y|\over(r^2+x^2)^{1/2}(r^2+y^2)^{1/2}}~.}
Since we are discussing constant values of~$D$ and~$\sigma$, we need to integrate the two-point functions in~\jjsthree\ over the sphere, and hence we will also need to understand possible contact terms at coincident points. Contact terms are short-distance contributions, which can be analyzed in flat space, and hence we can use results from section~2.

We begin by studying~$\d^2_D F \bigr|_{\CV = 0}$. Since~$\langle J(x) J(y)\rangle$ does not contain a contact term on dimensional grounds, we can calculate~$\d_D^2 F \bigr|_{\CV = 0}$ by integrating this two-point function over separated points on the sphere,
\eqn\Dsecder{{1\over r^4}{\d^2 F\over \d D^2} \bigg |_{\CV=0} = - {\tau_{ff} \over 16 \pi^2 r^4} \int_{S^3} \sqrt g \, d^3x \,  \int_{S^3} \sqrt g \, d^3y \;  {1\over s(x,y)^2} = - {\pi^2 \over 4} \, \tau_{ff} < 0~.}
The answer is finite and only depends on the constant~$\tau_{ff}$. The sign follows from unitarity.

The second derivative~$\d^2_\sigma F \bigr|_{\CV = 0}$ involves the integrated two-point function~$\langle K(x)K(y) \rangle_{S^3}$, which has a non-integrable singularity at coincident points. Since the resulting divergence is a short-distance effect, it can be understood in flat space. We can regulate the divergence by excising a small sphere of radius~$1\over \Lambda$ around~$x=y$.  Now the integral converges, but it leads to a contribution proportional to~$\Lambda$.  This contribution is canceled by a contact term~$\langle K(x) K(0)\rangle \sim \Lambda \delta^{(3)}(x - y)$.  The divergence and the associated contact term are related to the seagull term discussed in section~2. The removal of the divergence is unambiguously fixed by supersymmetry and current conservation, so that the answer is finite and well defined.\foot{To see this, note that in momentum space~$\langle J(p) J(-p)\rangle \sim {1 \over |p|}$. Supersymmetry implies that~$\langle K(p) K(-p) \rangle \sim p^2 \langle J(p) J(-p)\rangle \sim |p|$.  Thus, a contact term proportional to~$ \Lambda$ in $\langle K(p) K(-p) \rangle$ is incompatible with the two-point function of $J$ at separated points.  This shows that any UV cutoff that preserves supersymmetry does not allow a contact term, and hence it must lead to a finite and unambiguous answer for~$\int d^3 x \, \langle K(x) K(0) \rangle$. By contrast, excising a sphere of radius~$1\over \Lambda$ does not respect supersymmetry, and thus it requires a contact term.} This leads to
\eqn\sigmasecder{
{1\over r^2}{ \d^2 F \over \d \sigma^2}  \biggr|_{\CV=0} = - {\tau_{ff} \over 8\pi^2 r^2} \int_{S^3} \sqrt g \, d^3x \,  \int_{S^3}  \sqrt g \, d^3y \; {1\over s(x,y)^4} = {\pi^2\over 4} \, \tau_{ff} > 0~.}
Alternatively, we can evaluate the integral by analytic continuation of the exponent~$4$ in the denominator from a region in which the integral is convergent. Note that we have integrated a negative function to find a positive answer. This change of sign is not in conflict with unitarity, because we had to subtract the divergence.

Finally, the mixed derivative~$\d_D \d_\sigma F \bigr|_{\CV = 0}$ is obtained by integrating the two-point function~$\langle J(x) K(y) \rangle_{S^3}$, which vanishes at separated points. However, it may contain a non-vanishing contact term~\threefun, and hence it need not integrate to zero on the sphere. Such a contact term gives rise to
\eqn\mixedsdk{{1 \over r^3} {\d^2 F \over \d D \, \d \sigma} \, \biggr|_{\CV = 0} = -\pi \kappa_{ff}~.}
As we explained in section~2, the fractional part of~$\kappa_{ff}$ is a well-defined observable in the SCFT.

\newsec{Localization and~$F$-Maximization}

As we have explained in the introduction, localization embeds the SCFT of interest into the deep IR of an RG flow from a free theory in the UV. We can then compute~$F[\CV]$ on a three-sphere for certain supersymmetric choices of $\CV$,
\eqn\ssss{\sigma = m~, \qquad D = {i m \over r}~,}
with all other fields in~$\CV$ vanishing. Here~$m$ is a real constant that can be thought of as a real mass associated with the flavor symmetry that couples to~$\CV$. Hence~$D$ is imaginary. In order to place the theory on the sphere, we must choose an $R$-symmetry. As explained in~\refs{\JafferisUN,\HamaAV,\FestucciaWS}, the real parameter~$m$ can be extended to complex values,
\eqn\complexm{m \rightarrow m + {i t \over r}~,}
where~$t$ parameterizes the choice of~$R$-symmetry in the UV. The free energy computed via localization is then a holomorphic function of~$m + {i t \over r}$.

In general, the UV $R$-symmetry parametrized by~$t$ does not coincide with the superconformal~$R$-symmetry in the IR. This only happens for a special choice,~$t = t_*$. In this case~$F[m + {i t_* \over r}]$ encodes the free-energy and various current correlation functions in the SCFT on the sphere, exactly as in section~3. Expanding around~$m=0$, we write
\eqn\expandF{F\Big[m + {i t_* \over r}\Big]=F_0 + mr F_1 + \half (mr)^2 F_2+\cdots~.}

As we explained in section~3, the Chern-Simons term~\ggcs\ in the background gravity fields leads to complex~$F_0$, but it only affects the imaginary part. This explains the complex answers for~$F_0$ found in the localization computations of~\refs{\KapustinKZ\JafferisUN-\HamaAV}. Alternatively, we can remove the imaginary part by adding a Chern-Simons counterterm with appropriate fractional coefficient, at the expense of invariance under large background gauge transformations. The real part of~$F_0$ is not affected. It appears in the $F$-theorem~\fthm.

The first order term~$F_1$ arises because of the flavor-$R$ Chern-Simons term~\fgcs, which reduces to~\flavorgravity\ on the three-sphere. Restricting to the supersymmetric subspace~\ssss, we find that
\eqn\fone{F_1 = 2\pi \kappa_{fr}~.}
This accounts for the non-vanishing, real~$F_1$ found in~\refs{\KapustinKZ\JafferisUN-\HamaAV} and shows that~$\kappa_{fr}$ can be computed using localization. As we explained above, this term is not compatible with conformal symmetry. We can set it to zero and restore conformal invariance by adding an appropriate flavor-$R$ Chern-Simons counterterm, at the expense of invariance under large background gauge transformations.

The imaginary part of~$F_1$ always vanishes, in accord with conformal symmetry. Using holomorphy in~$m + {i t \over r}$, we thus find
\eqn\extr{{\d \over \d t} \Re F\biggr|_{m=0,  t = t_*}= -{1 \over r} {\d \over \d m} \Im F\biggr|_{m=0,  t = t_*}=0~.}
This is the condition proposed in~\JafferisUN.

The real part of~$F_2$ arises from~\Dsecder\ and~\sigmasecder,
\eqn\finalsectwo{\Re F_2 =  {1\over r^2}{\d^2\over \d m^2}  \Re F\biggr|_{m=0, t = t_*} =  {\pi^2\over2} \, \tau_{ff}~,}
while the imaginary part is due to the flavor-flavor Chern-Simons term~\ffcs.  Using~\mixedsdk, we obtain
\eqn\imsecder{\Im F_2 =  {1 \over r^2} {\d^2 \over \d m^2} \Im F \bigg|_{m = 0, t=t_*} =  -2 \pi \kappa_{ff}~.}
Combining the real and imaginary parts,
\eqn\Ftf{F_2 = {\pi^2 \over 2} \tau_{ff} -  2  \pi i \kappa_{ff}~.}
Thus, both~$\tau_{ff}$ and~$\kappa_{ff}$ are computable using localization.

If we denote by~$F(t) = F[0 + {i t \over r}]$ the free energy for~$m = 0$, we can summarize~\extr\ and~\finalsectwo\ as follows,
\eqn\fmaxonet{{\d \over \d t} \Re F \biggr|_{t = t_*} = 0~,
\qquad {\d^2 \over \d t^2} \Re F  \biggr|_{t = t_*} = -{\pi^2 \over 2} \, \tau_{ff} < 0~.}
The generalization to multiple Abelian flavor symmetries is straightforward and leads to~\fmax,
\eqn\fmaxii{{\d \over \d t^a} \Re F \biggr|_{t = t_*} = 0~,
\qquad {\d^2 \over \d t^a \d t^b} \Re F  \biggr|_{t = t_*} = -{\pi^2 \over 2} \, \tau_{a b}~,}
where the matrix~$\tau_{a b}$  is determined by the flat-space two-point functions of the Abelian flavor currents~$j^\mu_a$ at separated points,
\eqn\currtpii{\langle j^\mu_a (x) j_b^\nu (0)\rangle = {\tau_{a b} \over 16 \pi^2 } \left(\delta^{\mu\nu} \d^2 - \d^\mu \d^\nu\right) {1\over x^2}~.}
Unitarity implies that~$\tau_{ab}$ is a positive definite matrix. Note that our condition on the second derivatives is reminiscent of a similar condition in~\BarnesBM. However, the precise relation of~\BarnesBM\ to the three-sphere partition function is not understood.

As an immediate corollary, we obtain a non-renormalization theorem for the two-point function coefficients~$\tau_{ab}$ and $\kappa_{ab}$. Since localization sets all chiral fields to zero, the free energy is independent of all superpotential couplings, and hence all exactly marginal deformations. Thus~$\tau_{ab}$ and $\kappa_{ab}$ are independent of exactly marginal deformations.

We would briefly like to mention the connection of~\fmaxii\ to the~$F$-theorem~\fthm. It is analogous to the relationship between~$a$-maximization and the~$a$-theorem in four dimensions~\IntriligatorJJ. Since relevant deformations in the UV generally break some flavor symmetries, there are more flavor symmetries in the UV than in the IR. Maximizing over this larger set in the UV should result in a larger value of~$F$, thus establishing~\fthm. This simple argument applies to a wide variety of RG flows, but there are several caveats similar to those discussed in~\IntriligatorJJ. An important restriction is that the argument only applies to flows induced by superpotential deformations. For such flows, the free energy is the same function in the UV and in the IR, since it is independent of all superpotential couplings. One can say less about RG flows triggered by real mass terms,  since the free energy depends on them nontrivially.

One of the caveats emphasized in~\IntriligatorJJ\ is the existence of accidental symmetries in the IR of many RG flows. Similarly, to use localization at the point~$t = t_*$, we need to find an RG flow with an~$R$-symmetry that connects the SCFT in the~IR to a free theory in the UV. This is generally impossible if there are accidental symmetries in the~IR. Nevertheless, the maximization principle~\fmaxii\ holds. It would be interesting to find a three-dimensional analog of~\KutasovIY, which would enable exact computations in the presence of accidental symmetries. See~\AgarwalWD\ for recent work in this direction.

\newsec{Examples}

\subsec{Free Chiral Superfield}

Consider a free chiral superfield~$\Phi$ of charge~$+1$, coupled to~$\sigma$ and~$D$ in a background vector multiplet. The action on the sphere is given by
\eqn\ff{ S= \int_{S^3} \sqrt g \, d^3 x \, \left(  |\nabla \phi|^2- i\b \psi \gamma^\mu \nabla_\mu \psi+ \sigma^2|\phi|^2-D|\phi|^2 + i\sigma \b\psi\psi +{3\over 4 r^2} |\phi|^2 \right)~.}
For constant~$\sigma$ and~$D$, we can compute the partition function by performing the Gaussian functional integral over~$\phi$ and~$\psi$,
\eqn\totalsum{
F=\sum_{n=1}^\infty n^2\log\left(n^2-{1\over 4}+(\sigma^2-D) r^2 \right)-\sum_{n=1}^\infty n(n+1)\log\left((n+\half)^2 + (\sigma r)^2 \right)~.}
The two sums arise from the bosonic and the fermionic modes respectively. (The eigenvalues of the relevant differential operators on~$S^3$ can be found in~\KlebanovGS.) As expected, the leading divergence cancels due to supersymmetry, but there are lower-order divergences.

Instead of evaluating~\totalsum, we will calculate its derivative,
\eqn\Dder{\eqalign{
{1 \over r^2} {\d F \over \d D}~&= \sum^\infty_{n=1} {(\sigma^2-D) r^2 -{1\over4}\over n^2-{1\over 4}+(\sigma^2-D) r^2 }-\sum^\infty_{n=1} 1\cr
& ={\pi\over 2}\sqrt{(\sigma^2-D) r^2 -{1\over4}}\coth\left[\pi   \sqrt{(\sigma^2-D) r^2-{1\over4}} \right]~,}}
where we set~$\sum_n 1 \to -{1\over 2}$ by zeta function regularization.\foot{Equivalently, we can remove the divergence by an appropriate FI counterterm~\fiterm\ for the background vector multiplet.} Similarly, we find
\eqn\sigmaderi{\eqalign{
{1 \over r} {\d  F \over \d \sigma}=~& -\pi \sigma r \sqrt{(\sigma^2-D) r^2-{1\over4}}\coth\left[\pi   \sqrt{(\sigma^2-D) r^2-{1\over4}} \right] \cr
& +\pi\left((\sigma r)^2+{1\over 4}\right)   \tanh(\pi\sigma r)~.}}
Note that~\Dder\ and~\sigmaderi\ both vanish when~$\sigma=D=0$, as required by conformal invariance. The derivative of the free energy on the supersymmetric subspace~\ssss\ is given by
\eqn\onsubspace{{1 \over r} {\d F \over \d m}  = \pi\left(\half+imr\right)\tanh\left(\pi m r\right)~.}
This exactly matches the result obtained via localization~\refs{\JafferisUN,\HamaAV}.

We can also compute the second derivatives
\eqn\secondders{{1\over r^4} {\d^2 F \over \d D^2} \biggl |_{\sigma=D=0}=-{\pi^2\over 4}~,\qquad {1 \over r^2} {\d ^2  F \over \d \sigma^2} \biggl |_{\sigma=D=0}={\pi^2\over 4}~,}
and therefore,
\eqn\seconddersi{{1 \over r^2} {\d^2  \over \d m^2} \Re F \biggl |_{m=0}={\pi^2\over2}~.}
Since~$\tau_{ff} = 1$ for a free chiral superfield of charge~$+1$, these results are consistent with~\Dsecder, \sigmasecder, and \finalsectwo.

Finally, we discuss the mixed second derivatives,
\eqn\mixed{{1 \over r^3} {\d^2 F \over \d D \d \sigma} \bigg |_{\sigma = D = 0} = 0~, \qquad \lim_{\sigma r \rightarrow \pm \infty} \, {1 \over r^3}  {\d^2 F \over \d D \, \d \sigma} \bigg |_{D = 0} = \pm {\pi \over 2}~.}
Comparing with~\mixedsdk, we see that~$\kappa_{ff}$ vanishes in the UV theory. If we give the chiral superfield a real mass by turning on a non-zero value for~$\sigma$, the RG flow to the IR will generate a contact term~$\kappa_{ff} =- \half \,  {\rm sgn} (\sigma) $. This corresponds to the half-integer Chern-Simons term that arises when we integrate out a massive fermion~\refs{\RedlichKN, \RedlichDV}. Therefore the free energy is not invariant under all large gauge transformations of the background vector multiplet on arbitrary manifolds.  In order to preserve invariance under large gauge transformations, we must add a half-integer Chern-Simons term for the background gauge field to~\ff.

Note that the first derivative~\onsubspace\ has infinitely many zeros. By holomorphy, this means that~$F(t)$ has infinitely many extrema, even for a free chiral superfield. However, only one physically acceptable extremum is a local maximum. The~$F$-maximization principle may help resolve similar ambiguities in less trivial examples.

\subsec{Pure Chern-Simons Theory}

Consider a dynamical~$\CN=2$ Chern Simons theory with gauge group~$U(1)$ and integer level $k$,
\eqn\supcs{{k\over 4\pi}\left(i\ep^{\mu\nu\rho}A_\mu\del_\nu A_\rho-2\sigma D\right)+({\rm fermions})~.}
Here~$A_\mu$ denotes the dynamical gauge field rather than a background supergravity field. This theory has an Abelian flavor symmetry with topological current~$j^\mu={i \over 2\pi}\ep^{\mu\nu\rho}\del_\nu A_\rho$, whose correlation functions vanish at separated points. We can couple~$j^\mu$ to a background gauge field~$a_\mu$, which resides in a vector multiplet that also contains the bosons~$\sigma_a, D_a$, and we also add a background Chern-Simons term  for~$a_\mu$,
\eqn\supcsi{{1\over 2\pi}\left(i\ep^{\mu\nu\rho} a_\mu \del_\nu A_\rho-\sigma_a D-D_a \sigma \right)+{q\over 4\pi}\left(i\ep^{\mu\nu\rho}a_\mu\del_\nu a_\rho-2\sigma_a D_a\right)+({\rm fermions})~.}
Here $q$ is an integer. This example is discussed at length in~\future.

Naively integrating out~$A_\mu$ generates a Chern-Simons term for~$a_\mu$ with fractional coefficient
\eqn\fraccs{\kappa_{ff}=q-{1\over k}~.}
On the supersymmetric subspace~\ssss\ appropriate to the three-sphere,  this term evaluates to~$F_{ff} = -i\pi\kappa_{ff} (mr)^2$. We can compare it to the answer obtained via localization. Following~\KapustinKZ, we find that
\eqn\integral{e^{-F}=\int d (\sigma r) \, \exp(i\pi r^2 (k \sigma^2 + 2 \sigma m + q m^2) ) = {1\over \sqrt { |k|}}e^{i \, {\rm sgn} (k)  \pi/4}\exp(i\pi \kappa_{ff} (mr)^2)~.}
We see that the term in~$F$ proportional to~$m^2$ agrees with the flat-space calculation.

\subsec{SQED with a Chern-Simons Term}

Consider~$\CN=2$ SQED with an integer level~$k$ Chern-Simons term for the dynamical~$U(1)_v$ gauge field and~$N_f$ chiral flavor pairs~$Q_i, \t Q_{\t i} \; (i, \t i = 1, \ldots, N_f)$ that carry charge~$\pm 1$ under~$U(1)_v$. The theory also has a global~$U(1)_a$ flavor symmetry~$\CJ$ under which~$Q_i, \t Q_{\t i}$ all carry charge~$+1$. Here~$v$ and~$a$ stand for vector and axial respectively. The theory is invariant under charge conjugation, which flips the sign of the dynamical~$U(1)_v$ gauge field and interchanges~$Q_i \leftrightarrow \t Q_{\t i}$. In the IR, the theory flows to an SCFT, which is labeled by the integers~$k$ and~$N_f$.

In~\future, this model is analyzed in perturbation theory for~$k \gg 1$. Computing the appropriate two-point functions of the axial flavor current and the~$R$-current leads to
\eqn\cstermsi{\kappa_{ff} = {\pi^2 N_f \over 4 k} + \CO\left({1 \over k^3}\right)~,\qquad \kappa_{fr} = -{N_f \over 2k} + \CO\left({1 \over k^3}\right)~.}
We can now compare these flat-space calculations to the result obtained via localization~\refs{\JafferisUN, \HamaAV}. In the notation of~\expandF, we find
\eqn\locres{\eqalign{& F_0 = N_f \log 2 + \half \log |k| - {i \pi \over 4} \left({\rm sgn} (k) - {N_f \over k}\right) + \CO\left({1 \over k^2}\right)~,\cr
& F_1 = - { \pi N_f  \over k} + \CO\left({1 \over k^3}\right)~,\cr
& F_2 =   \pi^2N_f - {i \pi^3 N_f \over 2 k} + \CO\left({1 \over k^2}\right)~.}}
The real part of~$F_0$ is the conventional free energy for the SCFT in the IR. The imaginary part of~$F_0$ corresponds to~\ggcs, whose coefficient we will not discuss here. The first order term~$F_1$ exactly matches the contribution of the flavor-$R$ term as in~\fone, while the imaginary part vanishes to this order in~$1 \over k$. This is due to the fact that the mixing of the~$R$-current and the axial current only arises at~$\CO\left({1 \over k^2}\right)$. Likewise, the imaginary part of~$F_2$ is captured by the flavor-flavor term as in~\imsecder. Finally, the real part of~$F_2$ is in agreement with~\finalsectwo, since the two-point function coefficient of~$\CJ$ is given by~$\tau_{ff} = 2N_f + \CO\left({1 \over k^2}\right)$.

\subsec{A Theory with a Gravity Dual}

Equation~\fmax\ can be checked in~$\CN=2$ SCFTs with $AdS_4$ supergravity duals. The AdS/CFT correspondence~\refs{\MaldacenaRE\GubserBC-\WittenQJ} relates global symmetries of the boundary theory to gauge symmetries in the bulk. The boundary values~$a^a_\mu$ of the bulk gauge fields~$A^a_{\mu}$ act as background gauge fields for the global symmetry currents~$j_a^\mu$ on the boundary. The boundary free energy~$F[a]$ in the presence of these background fields is equal to the on-shell supergravity action computed with the boundary conditions~$A^a_{\mu}(x, z)|_{z=0}= a^a_{\mu}(x)$. The matrix~$\tau_{ab}$ defined by the two-point functions~\currtp\ of global currents on the boundary is proportional to the matrix~${1\over g^2_{ab}}$ of inverse gauge couplings that appears in the bulk Yang-Mills term~\FreedmanTZ.

Consider M-theory on~$AdS_4\times X_7$, where $X_7$ is a Sasaki-Einstein seven manifold. This background preserves~$\CN=2$ supersymmetry on the three-dimensional boundary. The isometries of~$X_7$ lead to~$AdS_4$ gauge fields upon KK reduction from 11-dimensional supergravity. Hence, they correspond to global symmetries of the dual SCFT$_3$. Given a set of Killing vectors~$K_a$ on~$X_7$ that are dual to the global symmetry currents~$j_a^{\mu}$, the matrix~$\tau_{ab}$ is given by~\refs{\BarnesBW}
\eqn\tauFromSugra{
\tau_{ab} = {32 \pi N^{3\over 2} \over 3\sqrt{6}  ({\rm Vol}(X_7))^{3\over 2}} \int G(K_a, K_b) \, vol(X_7)~.}
Here~$G$ is the Sasaki-Einstein metric on~$X_7$ and~$vol (X_7)$ is the corresponding volume form. There are~$N$ units of flux threading~$X_7$. We can use~\tauFromSugra\ to compute~$\tau_{ab}$ in the gravity dual and compare to the answer obtained via localization on the boundary, providing a check of~\fmax.

\ifig\fthree{Flavored conifold quiver dual to M2-branes on the cone over~$Q^{1,1,1}$.}
{\epsfxsize=1.5in \epsfbox{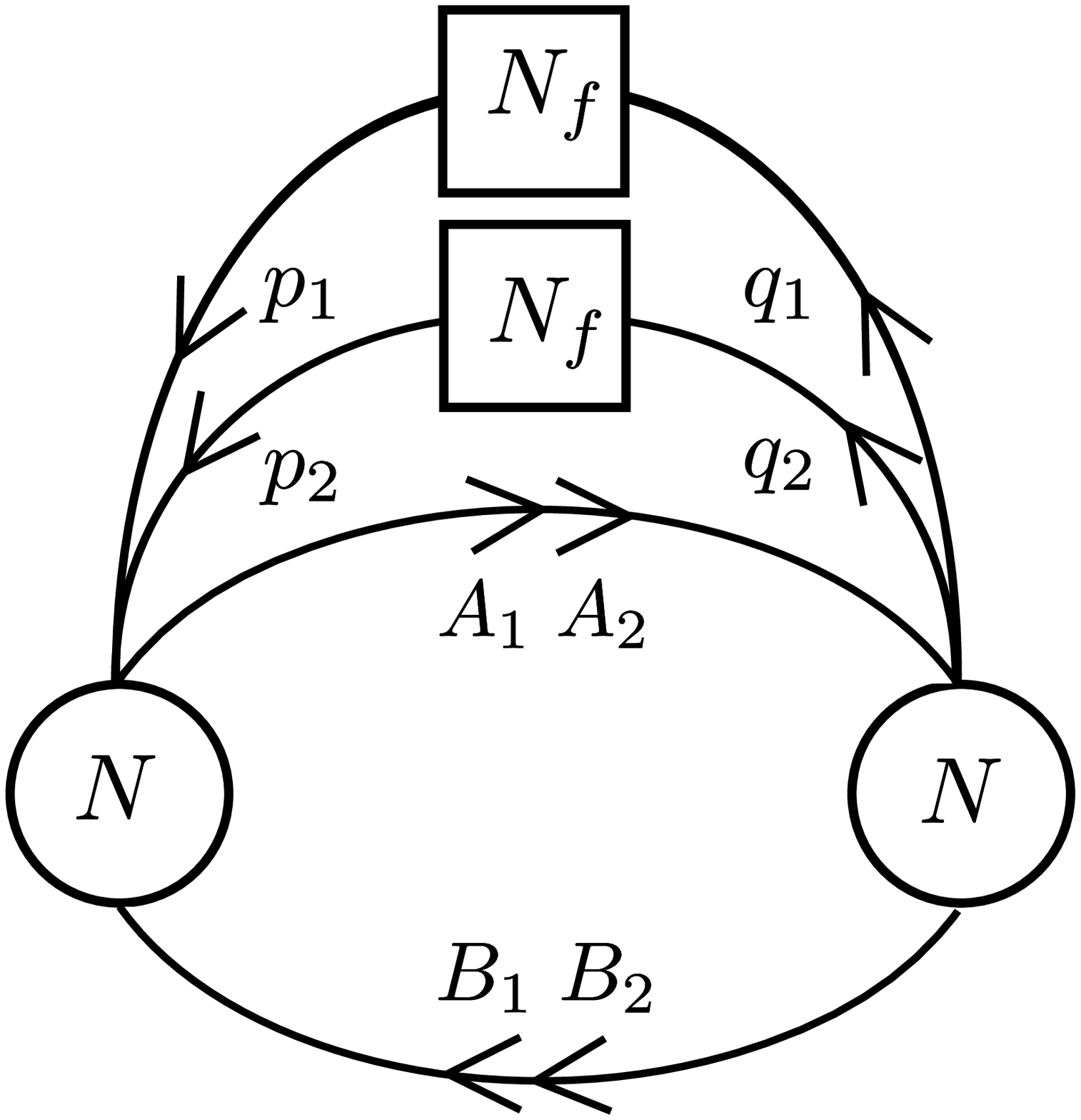} }

Consider, for instance, the theory depicted in~\fthree. It is the well-known conifold quiver with gauge group $U(N)\times U(N)$ and vanishing Chern-Simons levels, coupled to two~$U(N_f$) flavor groups. The superpotential is given by
\eqn\Wofquiver{
W= A_1B_1A_2B_2 - A_1B_2A_2B_1 + \sum_{l=1}^{N_f} \, \left( p_{1 l}A_1q_1^{l} +p_{2 l} A_2q_2^{l}\right)~.
}
This theory describes~$N$~M2 branes on a $\Z_{N_f}$ orbifold of the cone over~$Q^{1,1,1} \cong {SU(2)\times SU(2)\times SU(2) \over U(1)\times U(1)}$. It is expected to flow to the SCFT dual to~$AdS_4\times Q^{1,1,1}/\Z_{N_f}$ in the infrared~\refs{\BeniniQS,\JafferisTH}.

The large-$N$ partition function of this theory as a function of the trial~$R$-charges was computed in~\refs{\JafferisZI}.
For simplicity, we consider the free energy $F(t)$ as a function of a single mixing parameter $t$, which corresponds to the diagonal topological current\foot{The current is normalized so that certain diagonal monopole operators have charge~$\pm1$.}~$j^\mu~\!\!\!\sim~\!\!\!\ep^{\mu\nu\rho} \left(\Tr F^{(1)}_{\nu\rho} + \Tr F^{(2)}_{\nu\rho}\right)$. Here~$F^{(1)}_{\mu\nu}$ and~$F^{(2)}_{\mu\nu}$ are the field strengths of the two~$U(N)$ gauge groups. The function~$F(t)$ is maximized at~$t  =0$ and its second derivative is given by
\eqn\FofDmpp{
{ \d^2 F \over \d t^2} \biggl |_{t =0} = -{20 \pi\over 9\sqrt{3}} \left({N\over N_f}\right)^{3\over 2} \, .
}

We will now compute the two-point function coefficient~$\tau_{ff}$ of~$j^\mu$ via the AdS/CFT prescription~\tauFromSugra.
The Sasaki-Einstein metric on $Q^{1,1,1}$ takes the form
\eqn\Qooometric{
ds^2= {1\over 16} (d\psi +\sum_{i=1}^{3}\cos\theta_id\phi_i)^2 +{1\over 8}\sum_{i=1}^3 (d\theta^2_i+\sin^2{\theta_i}d\phi_i^2)~,
}
with~$\psi \in [0,4 \pi)$, $\phi_i \in [0, 2\pi)$, $\theta_i \in [0, \pi]$.
Using the results of \refs{\BeniniQS}, one can show that the Killing vector of~$Q^{1,1,1}$ that corresponds to the current~$j^\mu$ is given by\foot{This identification relies on a certain chiral ring relation involving two diagonal monopole operators, which was conjectured in~\refs{\BeniniQS,\JafferisTH}. Our final result below can be viewed as additional evidence for this conjecture.}
\eqn\KVm{
K = {1\over N_f}(-\del_{\phi_1}+\del_{\phi_2})~.}
Substituting into~\tauFromSugra\ and using~${\rm Vol}(Q^{1,1,1}/ \Z_{N_f})={\pi^4\over 8 N_f}$, we find
\eqn\tauFromSugramexp{
\tau_{ff} =  {40 \over 9\sqrt{3} \pi} \left({N\over N_f}\right)^{3\over 2}~.
}
Comparing~\FofDmpp\ and~\tauFromSugramexp, we find perfect agreement with~\fmax.

As was pointed out in~\refs{\JafferisZI,\MartelliQJ}, the~$F$-maximization principle is closely related to the volume minimization procedure of~\refs{\MartelliTP,\MartelliYB}. It is natural to conjecture that the two procedures are in fact identical. In other words, the two functions that are being extremized should be related, even away from their critical points. (A similar relation between~$a$-maximization in four dimensions and volume minimization was established in~\refs{\ButtiVN, \EagerYU}.) The example discussed above is consistent with this conjecture:~both the free energy at the critical point~\refs{\JafferisZI,\MartelliQJ} and its second derivative match.

\vskip 1cm

\noindent {\bf Acknowledgments:}
We would like to thank O.~Aharony, S.~Cremonesi, D.~Gaiotto, D.~Jafferis, A.~Kapustin, I.~Klebanov, J.~Maldacena, D.~Martelli, A.~Schwimmer, and J.~Sparks for many useful discussions. CC is a Feinberg postdoctoral fellow at the Weizmann Institute of Science. The work of TD was supported in part by a DOE Fellowship in High Energy Theory and a Centennial Fellowship from Princeton University. The work of GF was supported in part by NSF grant PHY-0969448.  TD and GF would like to thank the Weizmann Institute of Science for its kind hospitality during the completion of this project. ZK was supported by NSF grant PHY-0969448 and a research grant from Peter and Patricia Gruber Awards, as well as by the Israel Science Foundation under grant number~884/11.  The work of NS was supported in part by DOE grant DE-FG02-90ER40542. ZK and NS would like to thank the United States-Israel Binational Science Foundation (BSF) for support under grant number~2010/629.
Any opinions, findings, and conclusions or recommendations expressed in this
material are those of the authors and do not necessarily reflect the views of the funding agencies.

 \listrefs
 \end